\documentstyle[11pt]{article}
\makeatletter%
\def\nottoobig#1{{\hbox{$\left#1\vcenter to1.111\ht\strutbox{}\right.\n@space$}}}
\makeatother%

\makeatother%

\makeatletter%

\newcount\hour  \newcount\minutes  \hour=\time  \divide\hour by 60
\minutes=\hour  \multiply\minutes by -60  \advance\minutes by \time
\def\mmmddyyyy{\ifcase\month\or Jan\or Feb\or Mar\or Apr\or May\or Jun\or Jul\or
  Aug\or Sep\or Oct\or Nov\or Dec\fi \space\number\day, \number\year}
\def\hhmm{\ifnum\hour<10 0\fi\number\hour :%
  \ifnum\minutes<10 0\fi\number\minutes}
\def\Draft{{\it Draft of \mmmddyyyy}}

\def\ps@jtsheadings{%
\def\@oddhead{\it\rightmark\hfil\rm\thepage}%
\def\@oddfoot{\hfil\Draft}%
\if@twoside%
\def\@evenhead{\rm\thepage\hfil\it\leftmark}%
\def\@evenfoot{\Draft\hfil}%
\else
\let\@evenhead\@oddhead%
\let\@evenfoot\@oddfoot%
\fi%
}
\def\ps@jtsplain{%
\def\@oddhead{\hfil\Draft}%
\def\@oddfoot{\hfil\rm\thepage\hfil}%
\let\@evenfoot\@oddfoot%
\if@twoside \def\@evenhead{\Draft\hfil} \else \let\@evenhead\@oddhead \fi
}

\def\chaptermark#1{\markboth{\thechapter.\ #1}{\thechapter.\ #1}}%
\def\sectionmark#1{\markright{\thesection.\ #1}}

\def\section{\@startsection {section}{1}{\z@}
    {3.5ex plus1ex minus.2ex}{2.3ex plus.2ex}{\Large\bf}}
\def\subsection{\@startsection{subsection}{2}{\z@}
    {3.25ex plus1ex minus.2ex}{1.5ex plus.2ex}{\large\bf}}
\def\subsubsection{\@startsection{subsubsection}{3}{\z@}
    {3.25ex plus1ex minus.2ex}{1.5ex plus.2ex}{\normalsize\bf}}
\def\paragraph{\@startsection{paragraph}{4}{\z@}
    {3.25ex plus1ex minus.2ex}{1em}{\normalsize\bf}}
\def\subparagraph{\@startsection{subparagraph}{4}{\parindent}
    {3.25ex plus1ex minus.2ex}{1em}{\normalsize\bf}}

\makeatother%

\makeatletter \@beginparpenalty=10000 \makeatother

\def\underl#1 {\leavevmode\let\first=\relax\underli #1 }
\def\underli#1 {\ifx&#1\let\next=\relax\unskip
                \else\let\next=\underli\first\ulinebox{#1}\fi\let\first=\undersp\next}
\def\undersp{\penalty50\ulinebox{\space}\penalty50}
\def\ulinebox#1{\vtop{\hbox{\strut#1}\hrule}}%
\def\unice#1 {\underl #1 & }
\def\desclabel#1{\bf #1\hfil}
\def\desc{\list{}{%
\labelwidth=\leftmargin
\advance \labelwidth by -\labelsep
\let \makelabel=\desclabel}}

\makeatletter %
\newcommand{\Hat}[1]{ \widehat{#1} }

\newcommand{\implies}{\:\Rightarrow\:}

\newlength{\filength}
\newsavebox{\gcbox}
\sbox{\gcbox}{\framebox[\filength]{\rule{0ex}{2ex}}}

\newlength{\leftjustindent}
\newlength{\@leftjustindent}
\def\leftjust{\let\\\@leftjustcr\let\end\@endleftjust
  \addtolength{\@leftjustindent}{\leftjustindent}
  \vcenter\bgroup
  \halign\bgroup
    \hbox to\displaywidth{
      \rule{\@leftjustindent}{0ex}$\displaystyle##$\hfill
      }\crcr
}
\def\endleftjust{\crcr\egroup\egroup\endgroup}
\def\@endleftjust#1{\crcr\egroup\egroup\@checkend{#1}\endgroup}
\def\@leftjustcr{\crcr}

\newcommand{\red}[3]{ {  {\rm R}_{#1}^{#2}({#3}) }    }

\newtheorem{theorem}{Theorem}[section]
\newtheorem{corollary}[theorem]{Corollary}
\newcommand{\qedblob}{\mbox{\rule[-1.5pt]{5pt}{10.5pt}}}
\def\literalqed{{\ \nolinebreak\hfill\mbox{\qedblob\quad}}}

\def\qed{\literalqed}

\newtheorem{lemma}[theorem]{Lemma}

\newtheorem{proposition}[theorem]{Proposition}
\newcommand{\singlespacing}{\let\CS=
\@currsize\renewcommand{\baselinestretch}{1}\tiny\CS}
\newcommand{\singlespacingplus}{\let\CS=
\@currsize\renewcommand{\baselinestretch}{1.25}\tiny\CS}
\newcommand{\doublespacing}{\let\CS=
\@currsize\renewcommand{\baselinestretch}{1.75}\tiny\CS}
\newcommand{\draftspacing}{\let\CS=
\@currsize\renewcommand{\baselinestretch}{2.0}\tiny\CS}
\newcommand{\foospacing}{\let\CS=
\@currsize\renewcommand{\baselinestretch}{1.05}\tiny\CS}

\makeatother%

\mathcode`\0="0030      %
\mathcode`\1="0031
\mathcode`\2="0032
\mathcode`\3="0033
\mathcode`\4="0034
\mathcode`\5="0035
\mathcode`\6="0036
\mathcode`\7="0037
\mathcode`\8="0038
\mathcode`\9="0039
\singlespacingplus
\makeatletter
\clubpenalty=\@highpenalty
\widowpenalty=\@highpenalty
\makeatother

\makeatletter
\newcommand{\niceonespacing}{\let\CS=\@currsize\renewcommand{\baselinestretch}{1.1}\tiny\CS}\newcommand{\nicetwospacing}{\let\CS=\@currsize\renewcommand{\baselinestretch}{1.2}\tiny\CS}
\newcommand{\nicethreespacing}{\let\CS=\@currsize\renewcommand{\baselinestretch}{1.3}\tiny\CS}
\newcommand{\singlespacingplusplus}{\let\CS=\@currsize\renewcommand{\baselinestretch}{1.35}\tiny\CS}
\newcommand{\nicefourspacing}{\let\CS=\@currsize\renewcommand{\baselinestretch}{1.4}\tiny\CS}
\newcommand{\nicefivespacing}{\let\CS=\@currsize\renewcommand{\baselinestretch}{1.5}\tiny\CS}
\newcommand{\nicesixpacing}{\let\CS=\@currsize\renewcommand{\baselinestretch}{1.6}\tiny\CS}
\makeatother

\makeatletter
\def\@cite#1#2{[#1\if@tempswa , #2\fi]}
\makeatother

\makeatletter
\def\@citex[#1]#2{\if@filesw\immediate\write\@auxout{\string\citation{#2}}\fi
  \def\@citea{}\@cite{\@for\@citeb:=#2\do
    {\@citea\def\@citea{,\linebreak[0]}\@ifundefined
       {b@\@citeb}{{\bf ?}\@warning
       {Citation `\@citeb' on page \thepage \space undefined}}%
\hbox{\csname b@\@citeb\endcsname}}}{#1}}
\makeatother

\makeatletter
\def\ps@thesis{\def\@oddhead{\hfil\rm\thepage\hfil}\def\@oddfoot{}\def\@evenhead{\hfil\rm\thepage\hfil}\def\@evenfoot{}\def\chaptermark##1{}\def\sectionmark##1{}}
\makeatother

\makeatletter
\def\foobarpt{\textfont\z@\tenrm 
  \scriptfont\z@\ninrm \scriptscriptfont\z@\sevrm
\textfont\@ne\tenmi \scriptfont\@ne\ninmi \scriptscriptfont\@ne\sevmi
\textfont\tw@\tensy \scriptfont\tw@\ninsy \scriptscriptfont\tw@\sevsy
\textfont\thr@@\tenex \scriptfont\thr@@\tenex \scriptscriptfont\thr@@\tenex
\def\unboldmath{\everymath{}\everydisplay{}\@nomath\unboldmath
          \textfont\@ne\tenmi 
          \textfont\tw@\tensy \textfont\lyfam\tenly
          \@boldfalse}\@boldfalse
\def\boldmath{\@ifundefined{tenmib}{\global\font\tenmib\@mbi\@magscale1\global
        \font\tensyb\@mbsy \@magscale1\global\font
         \tenlyb\@lasyb\@magscale1\relax\@addfontinfo\@xiipt
              {\def\boldmath{\everymath
                {\mit}\everydisplay{\mit}\@prtct\@nomathbold
                \textfont\@ne\tenmib \textfont\tw@\tensyb 
                \textfont\lyfam\tenlyb\@prtct\@boldtrue}}}{}\@xiipt\boldmath}%
\def\prm{\fam\z@\tenrm}%
\def\pit{\fam\itfam\tenit}\textfont\itfam\tenit \scriptfont\itfam\ninit
   \scriptscriptfont\itfam\sevit
\def\psl{\fam\slfam\tensl}\textfont\slfam\tensl 
     \scriptfont\slfam\tensl \scriptscriptfont\slfam\tensl
\def\pbf{\fam\bffam\tenbf}\textfont\bffam\tenbf 
   \scriptfont\bffam\ninbf \scriptscriptfont\bffam\ninbf 
\def\ptt{\fam\ttfam\tentt}\textfont\ttfam\tentt
   \scriptfont\ttfam\nintt \scriptscriptfont\ttfam\nintt 
\def\psf{\fam\sffam\tensf}\textfont\sffam\tensf
    \scriptfont\sffam\tensf \scriptscriptfont\sffam\tensf
\def\psc{\@getfont\psc\scfam\@xiipt{\@mcsc\@magscale1}}%
\def\ly{\fam\lyfam\tenly}\textfont\lyfam\tenly 
   \scriptfont\lyfam\ninly \scriptscriptfont\lyfam\sevly
 \@setstrut \rm}

\makeatother

\newcommand{\p}{{\rm P}}

\newcommand{\np}{{\rm NP}}

\newcommand{\pnp}{{\p^{\rm NP}}}
\newcommand{\npnp}{{\np^{\rm NP}}}

\newcommand{\thetatwo}{{\Theta_2^p}}

\newcommand{\ph}{{\rm PH}}

\singlespacingplus

\def\pair#1{{{\langle\!\!~#1~\!\!\rangle}}}

\newcommand{\sigmastar}{\mbox{$\Sigma^\ast$}}

\newcommand{\calc}{\mbox{$\cal C$}}
\newcommand{\dr}{{\tt DodgsonRanking}}
\newcommand{\dw}{{\tt DodgsonWinner}}

\newcommand{\score}{{\mbox{\it{}Score}}}
\newcommand{\dsum}{{\mbox{\it{}DodgsonSum}}}
\newcommand{\ds}{{\tt DodgsonScore}}
\newcommand{\xthreec}{{\tt ExactCoverByThreeSets}}

\newcommand{\threedm}{{\tt ThreeDimensionalMatching}}
\newcommand{\merge}{\mbox{\it{}Merge}}
\newcommand{\mergeprime}{\mbox{{\it{}Merge}$'$}}

\newcommand{\voter}{\pair}

\newcommand{\bigo}{\mbox{$\cal O$}}
\newcommand{\condition}{\,\nottoobig{|}\:}

\foospacing
\title{%
Exact Analysis of Dodgson Elections:
Lewis Carroll's 1876 Voting System is Complete 
for Parallel Access to NP
}

\author{
Edith Hemaspaandra\thanks{%
Email: {\tt edith@bamboo.lemoyne.edu}.
Supported in part 
by grant
NSF-INT-9513368/\protect\linebreak[0]DAAD-315-PRO-fo-ab.  Work done in 
part while 
visiting 
Friedrich-Schiller-Universit\"at Jena and the 
University of Amsterdam.}
\\Department of Mathematics\\
Le Moyne College\\
Syracuse, NY 13214, USA
\and
Lane A. Hemaspaandra\thanks{%
Email: {\tt lane@cs.rochester.edu}.
Supported in part 
by grants NSF-CCR-9322513 and 
NSF-INT-9513368/\protect\linebreak[0]DAAD-315-PRO-fo-ab, and 
a University of Rochester Bridging Fellowship.  Work done in part while 
visiting 
Friedrich-Schiller-Universit\"at Jena and the 
University of Amsterdam.}
\\Department of Computer Science\\University of Rochester\\
            Rochester, NY 14627, USA
\and 
J\"{o}rg Rothe\thanks{%
Email: {\tt rothe@informatik.uni-jena.de}.
Supported in part 
by grant
NSF-INT-9513368/\protect\linebreak[0]DAAD-315-PRO-fo-ab
and a
NATO Postdoctoral Science Fellowship
from the Deut\-scher Aka\-de\-mi\-scher Aus\-tausch\-dienst
(``Ge\-mein\-sames Hoch\-schul\-sonder\-pro\-gramm~III
von Bund und L\"andern'').
Work done in part while visiting Le~Moyne College.}
\\Institut f\"ur Informatik\\
Friedrich-Schiller-Universit\"at Jena\\
07743 Jena, Germany}

\date{}

\makeatletter
\def\@listI{\leftmargin\leftmargini \parsep 4.5pt plus 1pt minus 1pt\topsep
6pt plus 2pt minus 2pt \itemsep  2pt plus 2pt minus 1pt}

\let\@listi\@listI
\@listi
\makeatother

\begin{document}

\typeout{WARNING:  BADNESS used to suppress reporting!  Beware!!}
\hbadness=3000%
\vbadness=10000 %

\bibliographystyle{alpha}

\pagestyle{empty}
\setcounter{page}{1}

\sloppy

\pagestyle{empty}
\setcounter{footnote}{0}

{\singlespacing

\maketitle

}

\begin{center}
{\large\bf Abstract}
\end{center}
\begin{quotation}
{\singlespacing

In 1876, Lewis Carroll proposed a voting system in which the winner is
the candidate who with the fewest changes in voters' preferences
becomes a Condorcet winner---a candidate who
beats all other candidates in pairwise 
majority-rule elections.  Bartholdi, 
Tovey, and Trick provided a
lower bound---NP-hardness---on the computational complexity of
determining the election winner in Carroll's system.  We provide a
stronger lower bound and an upper bound that matches our lower bound.
In particular, determining the winner in Carroll's system is complete
for parallel access to NP, i.e.,
it is complete for $\thetatwo$, for which it becomes the
most natural complete problem known.
It follows that determining the winner in Carroll's elections
is not NP-complete unless the polynomial hierarchy collapses.

}
\end{quotation}

\foospacing
\setcounter{page}{1}
\pagestyle{plain}
\sloppy
\section{Introduction}

The Condorcet criterion is that an election is won by any
candidate who defeats all others in pairwise majority-rule elections
(\cite{con:b:condorcet-paradox},
see~\cite{bla:b:polsci:committees-elections}).  The Condorcet Paradox,
dating from 1785~\cite{con:b:condorcet-paradox},
notes that not only is it not always the case
that Condorcet winners exist but, far worse,
when there are more than two candidates, pairwise majority-rule 
elections may yield strict cycles in the aggregate preference
even if each voter has non-cyclic 
preferences.\footnote{\protect\singlespacing{}The standard 
example is an election over candidates $a$, $b$, and $c$ in which
1/3 of the voters have preference $\pair{a<b<c}$,
1/3 of the voters have preference $\pair{b<c<a}$, and
1/3 of the voters have preference $\pair{c<a<b}$.  
In this case, though each voter individually has 
well-ordered preferences, the 
aggregate preference of the electorate is that $b$ trounces
$a$, $c$ trounces $b$, and $a$ trounces $c$.  In short, individually
well-ordered preferences do not 
necessarily aggregate to a well-ordered societal preference.}
This 
is a widely discussed and troubling feature of majority
rule (see, e.g., the discussion in~\cite{mue:b:public-choice-II}).

In 1876, Charles Lutwidge Dodgson---more commonly referred to today by
his pen name, Lewis Carroll---proposed an election system that is
inspired by the Condorcet 
criterion,\footnote{\protect\singlespacing{}Carroll
did not use this term.   Indeed, Black has 
shown that Carroll ``almost beyond a doubt'' was unfamiliar
with Condorcet's 
work~\protect\cite[p.~193--194]{bla:b:polsci:committees-elections}.} 
yet that sidesteps the
abovementioned problem~\cite{dod:unpubMAYBE:dodgson-voting-system}.
In particular, a Condorcet winner is a candidate who defeats each other
candidate in pairwise majority-rule elections.  In Carroll's system,
an election is won by the candidate who is ``closest'' to being a
Condorcet winner.  In particular, each candidate is given a score that
is the smallest number of exchanges of adjacent preferences in the
voters' preference orders needed to make the candidate a Condorcet
winner with respect to the resulting preference orders.  Whatever
candidate (or candidates, in the case of a tie) has the lowest score
is the winner.  This system admits ties but, as each candidate is
assigned an integer score, no strict-preference cycles are possible.

Bartholdi, Tovey, and Trick, in their paper ``Voting Schemes for which
It Can Be Difficult to Tell Who Won the
Election''~\cite{bar-tov-tri:j:who-won}, raise a  difficulty
regarding Carroll's election system.  Though the notion of winner(s)
in Carroll's election system is mathematically well-defined, Bartholdi
et al.~raise the issue of what the {\em computational complexity\/} is
of determining who is the winner.  Though most natural election
schemes admit obvious polynomial-time algorithms for determining who
won, in sharp contrast Bartholdi et al.~prove that Carroll's election
scheme has the disturbing property that it is NP-hard to determine
whether a given candidate has won a given election (a problem they dub
{\tt DodgsonWinner}), and that it is NP-hard even to determine whether
a given candidate has tied-or-defeated 
another given candidate (a problem they
dub {\tt DodgsonRanking}).   

Bartholdi, Tovey, 
and Trick's NP-hardness results
establish lower bounds for the complexity of 
{\tt DodgsonRanking} and {\tt DodgsonWinner}.
{\em We
optimally improve their two complexity lower bounds by proving that
both problems are hard for $\thetatwo$, the class of problems that can
be solved via parallel access to NP, and we provide matching upper
bounds.  Thus, we establish that both problems are
$\thetatwo$-complete.}  Bartholdi et al.~explicitly leave
open the issue of whether
\dr\ is NP-complete: ``...Thus \dr\ is as hard as an NP-complete
problem, but since we do not know whether \dr\ is in NP, we can say
only that it is NP-hard''~\cite[p.~161]{bar-tov-tri:j:who-won}.  
From our optimal lower bounds, it follows that 
neither \dw\ nor \dr\ is
NP-complete unless the polynomial hierarchy collapses.  

As to our proof method, in order to raise the known lower bound on
the complexity of Dodgson elections, we first study the ways in
which feasible algorithms can control Dodgson elections.  In
particular, we prove a series of lemmas showing how polynomial-time
algorithms can control oddness and evenness of election scores,
``sum'' over election scores, and merge elections.  These lemmas 
then lead to our hardness results.

We remark that it is somewhat curious finding 
``parallel access to NP''-complete 
(i.e., $\thetatwo$-complete)
problems that were introduced almost one hundred 
years before complexity theory 
itself existed.
In addition, \dw, which we prove
complete for this class, is extremely natural when
compared with previously known complete problems for this 
class, essentially all
of which have somewhat convoluted forms, e.g., asking whether a given
list of boolean formulas has the property that the number of
formulas in the list that are satisfiable is itself 
an odd number.
In contrast, the class NP, which is contained in $\thetatwo$,
has countless natural 
complete problems.  Also, we 
mention that Papadimitriou~\cite{pap:j:unique}
has shown that
{\tt{}Unique\-Optimal\-Traveling\-Salesperson}
is complete for $\p^{\rm NP}$, which contains $\thetatwo$.

\section{Preliminaries}\label{s:prelims}

In this section, we introduce some standard concepts and 
notations from computational complexity 
theory~\cite{pap:b:complexity,bov-cre:b:complexity,gar-joh:b:int}.
NP is the class of languages solvable in nondeterministic polynomial
time.  The polynomial
hierarchy~\cite{mey-sto:c:reg-exp-needs-exp-space,sto:j:poly}, PH, is
defined as $\ph = \p \cup \np \cup \np^{\rm NP} \cup \np^{\rm NP
^ {NP}} \cup \cdots$ where, for any class $\calc$, $\np^{\cal C}
= \bigcup_{C\in {\cal C}} \np^C$, and $\np^C$ is
the class of all languages that can be accepted by some NP machine
that is given a black box that in unit time answers
membership queries to $C$.  The polynomial hierarchy
is said to collapse if for some $k$ the $k$th term in the preceding
infinite union equals the entire infinite union.  Computer scientists
strongly suspect
that the polynomial hierarchy does not collapse, though proving
(or disproving) this remains a major open research issue.

The polynomial
hierarchy has a number of intermediate levels.  
The~$\thetatwo$ level of the polynomial hierarchy will
be of particular
interest to us.  $\thetatwo$,
which was 
first studied by 
Papadimitriou and Zachos~(\cite{pap-zac:c:two-remarks},
see also~\cite{wag:j:bounded}),
is 
the class of all languages that can be solved via $\bigo(\log n)$ 
queries to some NP set.
Equivalently, and more to the 
point for the purposes of this
paper, $\thetatwo$ equals the class of problems that can be solved
via parallel access to~NP~\cite{hem:j:sky,koe-sch-wag:j:diff}, as 
explained formally later in this section.
$\thetatwo$ falls between the first two
levels of the polynomial hierarchy: 
$\np \subseteq \thetatwo \subseteq \pnp \subseteq \npnp$.
During the past decade, $\thetatwo$ has 
played a quite active role in complexity theory.
Kadin~\cite{kad:j:pnplog} has proven that if NP
has a sparse Turing-complete set
then the polynomial hierarchy collapses to $\thetatwo$,
Hemachandra and Wechsung have shown that the question of
whether $\thetatwo$ and {\em sequential\/} access to~NP yield
the same class can be characterized in terms of 
Kolmogorov complexity~\cite{hem-wec:j:man-rand},
Wagner~\cite{wag:j:bounded} has shown that the
definition of $\thetatwo$ 
is extremely robust,
and Jenner and 
Tor\'{a}n~\cite{jen-tor:j:parallel} have shown that 
the robustness of the class 
$\thetatwo$ seems to fail for 
its function analogs.

Problems are encoded as languages of strings over some fixed alphabet
$\Sigma$ having at least two letters. $\sigmastar$ denotes the set of
all strings over~$\Sigma$. For any string $x \in \sigmastar$, let
$|x|$ denote the length of~$x$. 
For any set $A \subseteq \sigmastar$,
let $\overline{A}$ denote $\sigmastar \setminus A$.  
For any set $A \subseteq \sigmastar$,
let $||A||$ denote the cardinality of~$A$. For 
any multiset $A$, 
$||A||$ will denote the cardinality of $A$.  For example, 
if $A$ is the multiset containing one occurrence of the 
preference order $\pair{w<x<y}$ and seventeen occurrences of 
the preference order $\pair{w<y<x}$, then $||A|| = 18$.
As is standard, for each
language~$A \subseteq \sigmastar$ we use $\chi_{A}$ to denote
the characteristic function of~$A$, i.e., $\chi_{A}(x) = 1$ if $x \in A$
and $\chi_{A}(x) = 0$ if $x \not\in A$.
Let $\pair{\cdots}$ be any standard, multi-arity, easily
computable, easily invertible pairing function.
We will also use the notation $\pair{\cdots}$ to denote
preference orders, e.g., $\pair{w<x<y}$.  Which use is 
intended will be clear from context.  Whenever we speak
of a function that takes a {\em variable\/} number of arguments,
we will assume that the arguments, say $a_1, \ldots a_z$,
are encoded as $a_1\#\cdots\#a_z$, where~$\#$ is a symbol
not in the alphabet in which the arguments are encoded.  When
speaking of a variable-arity function being polynomial-time
computable, we mean that the function's running
time is polynomial
in $|a_1\#\cdots\#a_z| = z - 1 + |a_1| + \cdots + |a_z|$. 

In computational complexity theory, reductions are used to 
relate the complexity of problems.  Very informally, if $A$ reduces
to $B$ that means that, given $B$, one can solve $A$\@.  
For 
any $a$ and $b$ such that $\leq_a^b$ is a
defined reduction type, and any complexity class $\cal C$, let
$\red{a}{b}{{\cal C}}$ denote $\{ L \condition (\exists C \in {\cal
C})\,[ L \leq_a^b C]\}$.  We refer readers to the standard source,
Ladner, Lynch, and Selman~\cite{lad-lyn-sel:j:com}, for definitions
and discussion of the standard reductions.  However, we briefly and
informally present to the reader the definitions of the reductions
to be used in this paper.  $A \leq_m^p B$ (``$A$
polynomial-time many-one reduces to $B$'') if there is a
polynomial-time computable function $f$ such that $(\forall x \in
\sigmastar)\,[x\in A \iff f(x) \in B]$.  $A \leq_{tt}^p B$
(``$A$ polynomial-time truth-table reduces to $B$'') if there is a
polynomial-time Turing machine that, on input $x$, computes a 
query
that itself consists of a 
list of strings and, given that the machine
after writing the query is then 
given as its answer a list telling which of the
listed strings are in $B$, the machine then correctly determines
whether $x$ is in $A$ (this is not the original 
Ladner-Lynch-Selman definition, as we have
merged their querying machine and their evaluation machine, however this
formulation is common and equivalent).  Since a $\leq_{tt}^p$-reducing
machine, on a given input, asks all its questions in a {\em parallel\/}
(also called {\em non-adaptive\/}) manner,
the informal statement above that
$\thetatwo$ captures the complexity of ``parallel access to NP'' can
now be expressed formally as the claim $\thetatwo =
\red{tt}{p}{\np}$, which is 
known to hold~\cite{hem:j:sky,koe-sch-wag:j:diff}.  

As has become the norm, we always use hardness to denote hardness with
respect to $\leq_m^p$ reductions.  That is, for any class $\cal C$ and
any problem $A$, we say that $A$ is $\cal C${\em{}-hard\/} if 
$(\forall C \in
{\cal C}) [ C \leq_m^p A]$.  
For any class $\cal C$ and any problem~$A$, we say that $A$ is $\cal
C${\em{}-complete\/} if $A$ is $\cal C$-hard and $A \in {\cal C}$.
Completeness results are the standard method in computational 
complexity theory of categorizing the complexity of a problem, 
as a $\calc$-complete problem $A$ is both in $\calc$, and is 
the hardest problem in $\calc$ (in the sense that every 
problem in $\calc$ can be easily solved using $A$).

\section{The Complexity of Dodgson Elections}
\label{s:exact}

Lewis Carroll's voting
system~(\cite{dod:unpubMAYBE:dodgson-voting-system}, see
also~\cite{nie-rik:j:polsci:voting,bar-tov-tri:j:who-won}) works as
follows.  Each voter has strict preferences over the candidates.
Each candidate is assigned a score, namely, the smallest number of
sequential {\em exchanges of two adjacent candidates in the voters'
preference orders} (henceforward called ``switches'') needed to make
the given candidate a Condorcet winner.  We say that a candidate $c$
{\em ties-or-defeats\/} a candidate $d$ if the score of $d$ is not
less than that of $c$.  (Bartholdi et al.~\cite{bar-tov-tri:j:who-won}
use the term ``defeats'' to denote 
what we, for clarity, denote by ties-or-defeats;
though the notations are different, the sets being defined by Bartholdi
et al.~and in this paper are identical.)  A candidate $c$ is said to
win the Dodgson-type election if $c$ ties-or-defeats all other
candidates.  Of course, due to ties it is possible for two candidates
to tie-or-defeat each other, and so it is possible for more than one
candidate to be a winner of the election.

Recall that all preferences are assumed to be strict.
A candidate $c$ is a {\em Condorcet winner\/} 
(with respect to a given collection of voter preferences) 
if $c$ defeats (i.e., 
is preferred by
strictly more than half of the voters) 
each
other candidate
in pairwise majority-rule elections.
Of course, Condorcet winners do not necessarily exist for a given set
of preferences, but if a Condorcet winner does exist, it is unique.

We now return to Carroll's scoring notion to clarify
what is meant by the sequential nature of the 
switches, and to clarify by example that one switch changes
only one voter's preferences.
The ({\em
Dodgson\/}) {\em score\/} of any Condorcet winner is 0.  If a
candidate is not a Condorcet winner, but one switch (recall that a
switch is an exchange of two {\em adjacent\/} preferences in the
preference order of {\em one\/} voter) 
would make the candidate a Condorcet winner, then the candidate
has a score of 1.  If a candidate does not have a score of 0 or 1, but
two switches would make the candidate a Condorcet winner, then the
candidate has a score of 2.  Note that the two switches could both be
in the same voter's preferences, or could be one in one voter's
preferences and one in another voter's preferences.  Note also that
switches are sequential.  
For example, with two switches, one could
change a single voter's preferences from $\voter{a<b<c<d}$ to
$\voter{c<a<b<d}$,
where $e<f$ will denote the preference:
``$f$ is strictly preferred to $e$.''
With two switches, one could also change a single
voter's preferences from $\voter{a<b<c<d}$ to $\voter{b<a<d<c}$.  With
two switches (not one), one could also change two voters with initial
preferences of $\voter{a<b<c<d}$ and $\voter{a<b<c<d}$ to the new
preferences $\voter{b<a<c<d}$ and $\voter{b<a<c<d}$.  As noted 
earlier in this section, Dodgson 
scores of 3, 4, etc., are defined analogously, i.e.,
the Dodgson score of a candidate is
the smallest number of
sequential switches
needed to make
the given candidate a Condorcet winner.  (We note in passing
that Dodgson was before his time in more ways than one.  His definition
is closely related to an important concept that is now known in computer
science as ``edit-distance''---the minimum number of operations
(from some specified set of operations) required to transform one
string into another.  Though Carroll's single ``switch'' operation 
is not the richer set of operations most commonly used 
today when doing string-to-string 
editing (see, e.g.,~\cite{kru-san:b:sequence-comparison}), 
it does form a valid basis operation for transforming
between permutations, which after all are what preferences
are.)

Bartholdi et al.~\cite{bar-tov-tri:j:who-won}
define a number of decision problems
related to Carroll's system.  They prove that given preference lists,
and a candidate, and a number $k$, it is NP-complete to determine 
whether the
candidate's score is at most $k$ in the election specified by
the preference lists (they call this problem {\tt DodgsonScore}).
They define the problem \dr\ to be the problem of determining, given
preference lists and the names of two voters, $c$ and $d$, whether $c$
ties-or-defeats $d$.  They prove that this problem is NP-hard.  They also
prove that, given a candidate and preference lists, it is NP-hard to
determine whether the candidate is a winner of the election.

For the formal definitions of these three decision problems, 
a preference order is
strict (i.e., irreflexive and antisymmetric), 
transitive, and complete. Since we will freely identify voters with their
preference orders, and two different voters can have the same preference order, we
define a set of voters as a 
multiset of preference orders.

We will say that $\pair{C,c,V}$ is a {\em Dodgson triple\/} if $C$ is a set of
candidates, $c$ is a member of $C$, and $V$ is a
multiset of preference orders on $C$.  Throughout this
paper, we assume that, as inputs, multisets are coded as lists, i.e.,
if there are $m$ voters in the voter set
then $V = \pair{P_1,P_2,\ldots,P_m}$,
where $P_i$ is the preference order of the $i$th voter.
$\score(\pair{C,c,V})$
will denote the Dodgson score of $c$ in the 
vote specified by $C$ and $V$\@.  If $X$ is a decision problem,
then when we speak of an {\em instance\/} of $X$ we mean a 
string that satisfies the
syntactic conditions listed in 
the ``Instance'' field 
of the problem's definition 
(or implicit in that field in order 
for the problem to be syntactically
well-formed---e.g., preference 
lists must be over the right number and right set of candidates).
As is standard, since all such syntactic conditions
in our decision problems  are trivially checkable 
in deterministic polynomial time, this is equivalent to the 
language definitions that are also common;  in particular,
the language corresponding to decision problem $X$ is the set
$\{ x  \condition x$ is an instance of $X$, and the 
``Question'' of decision problem $X$ has the answer ``yes''
for~$x\}$.  Since reductions map between sets, whenever 
speaking of or constructing reductions we use this latter 
formalism.  

{\singlespacing

\begin{description}
\item[Decision Problem: ] \ds
\item[Instance: ] A Dodgson triple $\pair{C,c,V}$; a positive integer $k$.
\item[Question: ] Is $\score(\pair{C,c,V})$, the Dodgson score of candidate $c$
in the election specified by~$\pair{C,V}$,
less than or equal to $k$? 
\end{description}

\smallskip

\begin{description}
\item[Decision Problem: ] \dr
\item[Instance: ] A set of candidates $C$; two distinguished
members of $C$, $c$ and $d$;
a multiset $V$ of preference orders on $C$ (encoded as a list, as 
discussed above).
\item[Question: ] Does $c$ tie-or-defeat $d$ in the election? That is, is
$\score(\pair{C,c,V}) \leq \score(\pair{C,d,V})$?
\end{description}

\smallskip

\begin{description}
\item[Decision Problem: ] \dw
\item[Instance: ] A Dodgson 
triple $\pair{C,c,V}$.
\item[Question: ] Is $c$ a winner of the election?
That is, does $c$ tie-or-defeat all other candidates in the election?
\end{description}

} %

We now state the complexity of \dr.

\begin{theorem}\label{t:theta}
\dr\ is  $\thetatwo$-complete.
\end{theorem}

It follows immediately---since 
(a)~$\thetatwo = \np \implies \ph = \np$, and 
(b)~$\red{m}{p}{\np}=\np$---that \dr, though known 
to be NP-hard~\cite{bar-tov-tri:j:who-won}, cannot
be NP-complete 
unless the polynomial hierarchy collapses quite dramatically.

\begin{corollary}
If \dr\ is $\np$-complete, then $\ph = \np$.
\end{corollary}

Most of the rest of the paper is devoted to working towards a proof of
Theorem~\ref{t:theta}.  Wagner has provided a useful tool for proving
$\thetatwo$-hardness, and we state his result below as
Lemma~\ref{l:klaus}.  However, to be able to exploit this tool we
must explore the structure of Dodgson elections.  
In particular, we have to learn how 
to control oddness 
and evenness of election scores, how to add election scores, and 
how to merge
elections.
We do so
as
Lemmas~\ref{l:reduction}, \ref{l:SUMMATION}, and~\ref{l:merge},
respectively.  On our way towards a proof of
Theorem~\ref{t:theta}, using Lemmas~\ref{l:klaus}, \ref{l:reduction}, and
\ref{l:SUMMATION} we will first establish $\thetatwo$-hardness of a
special problem that is closely related to \dr. This result is
stated as Lemma~\ref{l:less} below.   It is not hard to prove
Theorem~\ref{t:theta} using Lemma~\ref{l:less} and Lemma~\ref{l:merge}.
Note that Lemma~\ref{l:merge} gives more than is needed merely to establish
Theorem~\ref{t:theta}. In fact, the way this lemma is stated even 
suffices to provide---jointly with Lemma~\ref{l:less}---a direct
proof of the
$\thetatwo$-hardness of \dw.

\begin{lemma}
\label{l:klaus}
{\bf \cite{wag:j:more-on-bh}}~
Let $A$ be some $\np$-complete set, and let $B$ be any set. If
there exists a polynomial-time computable function $g$ such that,
for all $k \geq 1$ and all 
strings $x_1, \ldots, x_{2k} \in \sigmastar$ satisfying
$\chi_A(x_1) \geq \chi_A(x_2) \geq \cdots \geq \chi_A(x_{2k})$, it 
holds that
\[||\{i\condition x_i \in A\}|| \mbox{ is odd }
\Longleftrightarrow g(x_1, \ldots,
x_{2k}) \in B,\]
then $B$ is $\thetatwo$-hard.\footnote{\protect\label{f:arity}
Recall 
the comments/conventions
of Section~\ref{s:prelims} regarding the handling 
of the arguments of variable-arity functions.  Wagner did not 
discuss this issue, but we note that his proof remains valid under 
the conventions 
of Section~\ref{s:prelims}.  These conventions
have been adopted as they shield Wagner's theorem from
a pathological type of counterexample 
(involving large, variable numbers of length zero 
inputs~($\epsilon$) followed by one other constant-length string)
noted by a referee
that, without the conventions, could render Wagner's theorem
true but never applicable.

Another difference in our statement of the theorem 
relative to Wagner's is that 
though we state the theorem for the class~$\thetatwo$, Wagner
used the class ``$\p_{\rm bf}^{\rm NP}$.''  However, this 
is legal as 
$\p_{\rm bf}^{\rm NP}$ is now
known to be equal to~$\thetatwo$ 
(see the discussion in~\cite[Footnote~1]{koe-sch-wag:j:diff}).}

\end{lemma}

\begin{lemma}
\label{l:reduction}
There exists an $\np$-complete set $A$ and a polynomial-time computable
function $f$ that reduces $A$ to \ds\ in such a way that, for every $x \in
\sigmastar$, 
$f(x) = \pair{\pair{C,c,V},k}$
is an instance of \ds\ with an odd number of voters
and
\begin{enumerate}
\item if $x\in A$ then $\score(\pair{C,c,V}) = k$, and
\item if $x\not \in A$ then $\score(\pair{C,c,V}) = k+1$.
\end{enumerate}
\end{lemma}

\begin{lemma}
\label{l:SUMMATION}
There exists a polynomial-time computable function $\dsum$
such that, for all $k$ and for all
$\pair{C_1,c_1,V_1}$, 
$\pair{C_2,c_2,V_2}$,~$\ldots$,
$\pair{C_k,c_k,V_k}$ 
satisfying $(\forall j)[||V_j||$ is odd$]$,
it holds that 
$$\dsum( \pair{  \,\pair{C_1,c_1,V_1}\,,\,
\pair{C_2,c_2,V_2},\, \ldots\,,\,
\pair{C_k,c_k,V_k}\,})$$
is a
Dodgson triple
having an odd number of
voters and such that 
\[ \sum_j \score(\pair{C_j,c_j,V_j}) =
\score(   
\dsum( \, \pair{  \,\pair{C_1,c_1,V_1}\,,\,
\pair{C_2,c_2,V_2},\, \ldots\,,\,
\pair{C_k,c_k,V_k}\,} \,) \, ).\]
\end{lemma}

Lemma~\ref{l:klaus}, Lemma~\ref{l:reduction}, and
Lemma~\ref{l:SUMMATION} together establish the $\thetatwo$-hardness of
a special problem that is closely related to the problems that we
are interested in, \dr\ and \dw.  Let us define the decision
problem {\tt TwoElectionRanking} ({\tt{}2ER}).

\smallskip

{\singlespacing

\begin{description}
\item[Decision Problem: ] {\tt TwoElectionRanking} ({\tt{}2ER}) 
\item[Instance: ] A pair of Dodgson triples 
$\pair{\pair{C,c,V},\pair{D,d,W}}$ both having an odd number of voters
and such that $c \neq d$.
\item[Question: ] Is $\score(\pair{C,c,V}) \leq \score(\pair{D,d,W})$?
\end{description}

} %

\begin{lemma}
\label{l:less}
{\tt TwoElectionRanking} is $\thetatwo$-hard.
\end{lemma}

We note in passing that {\tt{}2ER} 
is in 
$\red{tt}{p}{\np}$.   This fact
follows by essentially the same argument 
that will be used in the 
proof of Theorem~\ref{t:theta} to 
establish that theorem's upper bound.
Thus, since
$\thetatwo =
\red{tt}{p}{\np}$, we have---in light of 
Lemma~\ref{l:less}---that {\tt{}2ER} 
is $\thetatwo$-complete.  We also note in passing that,
since one can trivially rename candidates, {\tt{}2ER}
remains $\thetatwo$-complete in the variant in
which ``and such that $c\neq d$'' is removed
from the problem's definition.

In order to make the results obtained so far applicable to \dr\ and
\dw, we need the following lemma that tells us how to merge two
elections into a single election in a controlled manner. 

\begin{lemma}
\label{l:merge}
There exist polynomial-time computable functions $\merge$ 
and $\mergeprime$ such that, for
all Dodgson triples $\pair{C,c,V}$ and $\pair{D,d,W}$ 
for which \mbox{$c\neq d$} 
and both $V$ and $W$ represent odd numbers of
voters,
there exist $\widehat{C}$ and $\widehat{V}$ 
such that 
\begin{description}
\item[(i)] $\merge(\pair{C,c,V}, \pair{D,d,W})$ is an 
instance of \dr\ and
$\mergeprime(\pair{C,c,V}, \pair{D,d,W})$ is an 
instance of \dw, 
\item[(ii)] $\merge(\pair{C,c,V}, \pair{D,d,W})
= \pair{\widehat{C},c,d,\widehat{V}}$ and \\
$\mergeprime(\pair{C,c,V}, \pair{D,d,W})
= \pair{\widehat{C},c,\widehat{V}}$, 
\item[(iii)] $\score(\pair{\widehat{C},c,\widehat{V}}) = 
\score(\pair{C,c,V}) + 1$,
\item[(iv)] $\score(\pair{\widehat{C},d,\widehat{V}}) = 
\score(\pair{D,d,W}) + 1$,
and  
\item[(v)] for each $e \in \widehat{C}\setminus \{c,d\}$,
$\score(\pair{\widehat{C},c,\widehat{V}}) <
\score(\pair{\widehat{C},e,\widehat{V}})$.
\end{description}
\end{lemma}

We now prove these lemmas.

\medskip

\noindent {\bf Proof of Lemma~\ref{l:reduction}.} 
Bartholdi et al.~\cite{bar-tov-tri:j:who-won}
prove the NP-hardness of
\ds\ by reducing \xthreec\ to it. However, their
reduction doesn't have the additional 
properties that we need in this lemma.
We will construct a reduction from the NP-complete problem \threedm\
({\tt{}3DM})~\cite{gar-joh:b:int} to \ds\ that {\em does\/} have the 
additional properties we need.  
Let us first give the definition of {\tt{}3DM}:

{\singlespacing

\begin{description}
\item[Decision Problem: ] \threedm\ ({\tt{}3DM})
\item[Instance: ] Sets $M$, $W$, $X$, and $Y$,
where $M \subseteq W \times X \times Y$
and $W$, $X$, and
$Y$ are disjoint, nonempty sets having the same number of elements.
\item[Question: ] Does $M$ contain a {\em matching}, 
i.e., a subset $M' \subseteq
M$ such that $||M'|| = ||W||$ and no two elements of $M'$ agree in any
coordinate?
\end{description}

}%

We now describe a polynomial-time reduction $f$ (from {\tt{}3DM} to
{\tt{}DodgsonScore}) having the desired properties.  Our reduction is
defined by $f(x) = f'(f''(x))$, where $f'$ and $f''$ are as described
below.  Informally, $f''$ turns all inputs into a standard format
(instances of {\tt{}3DM} having $||M|| >1$), and $f'$ assumes its
input has this format and implements the actual reduction.

Let $f''$ be a polynomial-time function that has 
the following properties.  
\begin{enumerate}
\item If $x$ is not an instance of {\tt{}3DM} or is 
an instance of {\tt{}3DM} having $||M|| \leq 1$, then 
$f''(x)$ will output an instance $y$ of 
{\tt{}3DM} for which $||M|| > 1$ and, furthermore,
it will hold that $y \in \mbox{{\tt{}3DM}} \iff x \in
\mbox{{\tt{}3DM}}$.
\item If $x$ is 
an instance of {\tt{}3DM} having $||M|| > 1$, then 
$f''(x) = x$.
\end{enumerate}
It is clear that such functions 
exist.  In particular,
for concreteness, 
let $f''(x)$ be
$\pair{ \{(d,e,p), (d,e,p')\} , 
\{d,d'\}, \{e,e'\},\{p,p'\} }$
if 
$x$ is not an instance of {\tt{}3DM} or both
$x \not\in \mbox{{\tt{}3DM}}$ and $x$ is 
an instance of {\tt{}3DM} having $||M|| \leq 1$;
let $f''(x)$ be 
$\pair{ \{(d,e,p), (d',e',p')\} , 
\{d,d'\}, \{e,e'\},\{p,p'\} }$
if 
$x$ is 
an instance of {\tt{}3DM} having $||M|| \leq 1$
and such that $x\in \mbox{{\tt{}3DM}}$;
let $f''(x)$ be $x$ otherwise.

We now describe $f'$.  Let $x$ be our input.  If $x$ is not 
an instance of {\tt{}3DM} for which $||M|| > 1$ then 
$f'(x) = 0$;  this is just for definiteness, as due to 
$f''$, the only actions of $f'$ that matter are when the 
input is 
an instance of {\tt{}3DM} for which $||M|| > 1$.
So, 
suppose $x=\pair{M,W,X,Y}$ is an instance of {\tt{}3DM}
for which $||M|| > 1$.  Let $q=||W||$.  
Define $f'(\pair{M,W,X,Y}) = \pair{\pair{C,c,V},3q}$ as follows:
Let $c$, $s$, and $t$ be elements
not in $W \cup X \cup Y$.
Let $C = W \cup X \cup Y \cup \{c,s,t\}$
and let $V$ consist of the following two subparts:
\begin{enumerate}
\item Voters simulating elements of $M$.
Suppose the elements of $M$ are enumerated as $\{(w_i,x_i,y_i) \condition
1 \leq i \leq ||M||\}$.  (The $w_i$ are not intended to be 
an enumeration of $W$.  Rather, they take on values from $W$ 
as specified by $M$.  In particular, $w_j$ may equal $w_k$ even
if $j\neq k$.  The
analogous comments apply to the $x_i$ and $y_i$ variables.)
For every triple $(w_i,x_i,y_i)$ in $M$, we will create a voter.
If $i$ is odd, we create the voter
$\voter{s < c < w_i < x_i < y_i < t < \cdots }$, where the elements after
$t$ are the elements of $C \setminus 
\{s,c,w_i,x_i,y_i,t\}$ 
in arbitrary order.
If $i$ is even, we do the same, except that we exchange $s$ and $t$.
That is, we create the voter
$\pair{t < c < w_i < x_i < y_i < s < \cdots }$, where the elements after
$s$ are 
the elements of $C \setminus 
\{s,c,w_i,x_i,y_i,t\}$ 
in arbitrary order.

\item $||M||-1$ voters who prefer $c$ to all other candidates.
\end{enumerate}

We will now show that $f$ has the desired properties.
It is immediately clear that $f''$ and $f'$, and thus 
$f$, are polynomial-time computable.  It is also clear from
our construction that, for each $x$, $f(x)$ is an
instance of \ds\ having an odd number of
voters
since,
for every instance 
$\pair{M,W,X,Y}$ of {\tt{}3DM} with $||M||>1$, 
$f'(\pair{M,W,X,Y})$ is an instance of \ds\ with 
$||M|| + (||M||-1)$ voters, and since $f''$ always 
outputs instances of this form.
It remains to show that, for every 
instance 
$\pair{M,W,X,Y}$ of {\tt{}3DM} with $||M||>1$:
\begin{description}
\item[(a)] if $M$ contains a matching, then $\score(\pair{C,c,V}) = 3q$, and
\item[(b)] if $M$ does not contain a matching, then 
$\score(\pair{C,c,V}) = 3q+1$.
\end{description}
Note that if we prove this, it is clear that $f$ has the 
properties~(1) and~(2) of Lemma~\ref{l:reduction}, in light
of the properties of
$f''$.
Note that, recalling that we may 
now assume that $||M||>1$, by construction
$c$ is preferred to $s$ and $t$ by more than half of the
voters, and is preferred to all other candidates by $||M||-1$ of the  
$2||M|| - 1$ voters.

Now suppose that $M$ contains a matching $M'$.
Then $||M'|| = q$, and every element in $W \cup X \cup Y$ occurs in $M'$.
$3q$ switches turn $c$ into a Condorcet winner as follows.
For every element $(w_i,x_i,y_i) \in M'$, switch $c$ upwards 3 times
in the voter corresponding to $(w_i,x_i,y_i)$.
For example, if $i$ is odd, this voter changes from
$\voter{s < c < w_i < x_i < y_i < t < \cdots }$ to
$\voter{s < w_i < x_i < y_i < c < t < \cdots }$.
Let $z$ be an arbitrary element of $W \cup X \cup Y$.
Since $z$  occurs in $M'$,
$c$ has gained one vote over~$z$. Thus, $c$ is preferred to $z$ by
$||M||$ of the $2||M||-1$ voters. Since $z$ was arbitrary, 
$c$ is a Condorcet winner.

On the other hand,
$c$'s  Dodgson score can never be less than $3q$, because
to turn $c$ into a Condorcet winner,
$c$ needs to gain one vote over $z$
for every $z \in W \cup X \cup Y$.
Since $c$ can gain only {\em one\/} vote over {\em one\/} candidate for each switch,
we need at least $3q$ switches to turn $c$ into a Condorcet winner.
This proves condition~(a).

To prove condition~(b), first note that 
there is a ``trivial'' way to turn $c$ into a Condorcet winner with
$3q+1$ switches: Just switch $c$ to the top of the preference order of the
first voter. The first voter was of the form
$\voter{s < c < w_1 < x_1 < y_1 < t < \cdots }$, where the elements after $t$
are exactly all elements
in $W \cup X \cup Y \setminus \{w_1, x_1, y_1\}$, in arbitrary order. 
Switching $c$ upwards $3q+1$ times moves $c$ to the top of the preference order
for this voter, and gains one vote for $c$ over all candidates in $W \cup X
\cup Y$, which turns $c$ into a Condorcet winner. This shows that
$\score(\pair{C,c,V}) 
\leq 3q+1$, regardless of whether $M$ has a matching or not.

Finally, note that a Dodgson score of $3q$
implies that $M$ has a matching.
As before, every switch has to involve $c$ and an element of $W \cup X \cup Y$.
(This is because $c$ must gain a vote over $3q$ other candidates---$W 
\cup X \cup Y$---and so any switch involving $s$ or $t$ would 
ensure that at most $3q-1$ switches were available for gaining 
against the $3q$ members of $W \cup X \cup Y$, thus 
ensuring failure.)
Thus, for every voter, $c$ switches at most three times to become a
Condorcet winner.
Since $c$ has to gain one vote in particular over each element in 
$Y$, and to ``reach'' an element in $Y$ it must hold that $c$
first switches over the elements of $W$ and $X$ that due to 
our construction fall between it and the 
nearest $y$ element (among the $||M||$ voters simulating 
elements of $M$---it is clear that if any switch involves at least
one of
the $||M||-1$ dummy voters this could never lead to a Dodgson
score of $3q$ for $c$),
it must be the case that $c$ switches upwards exactly three times for
exactly $q$ voters corresponding to elements of $M$.
This implies that the $q$ elements of $M$ that correspond to these $q$ voters
form a matching, thus proving condition~(b).~\qed

\medskip

\noindent {\bf Proof of Lemma~\ref{l:SUMMATION}.}
We define $$\dsum( \pair{  \,\pair{C_1,c_1,V_1}\,,\,
\pair{C_2,c_2,V_2},\, \ldots\,,\,
\pair{C_k,c_k,V_k}\,}) = \pair{\Hat{C},c,\Hat{V}},$$
where $\Hat{C}$, $c$, and $\Hat{V}$ will be as
constructed in this proof.

Let $c = c_1$.
Without loss of generality (by renaming if needed), 
we assume that 
$c_1 = c_2 = \cdots = c_k$, and that 
$(\forall i,j)[i\neq j \implies C_i \cap C_j = \{c\}]$.

Also, for each $i$, 
enumerate $C_i \setminus \{c\}$ as $\{c_{i,1},c_{i,2}, \ldots, 
c_{i,||C_i||-1}\}$.  To make our preference orders easier 
to read, whenever in a preference order we write in the text
``$\overrightarrow{C_i}$,'' this should be viewed as being replaced by 
the text string ``$
c_{i,1} < c_{i,2} < \cdots <
c_{i,||C_i||-1}$.'' 

As our candidate set, we will take all the old candidates 
from the given elections, i.e., $\{
c,
c_{1,1},c_{1,2},\ldots,c_{1,||C_1||-1},
c_{2,1},c_{2,2},\ldots,c_{2,||C_2||-1},
\cdots,
c_{k,1},c_{k,2},\ldots,c_{k,||C_k||-1}\}$, plus a
set
$S$ of new ``separator'' candidates, whose only purpose is 
to avoid interference.
We will ensure that 
$c$ is preferred to all elements of $S$ by a majority of
the voters.

Formally, let $S = \{s_i \condition 1 
\leq i \leq \sum_j ||C_j||\cdot ||V_j||\}$, and
let $\Hat{C} = S \cup \bigcup_j C_j$.
As a notational convenience,
whenever in a preference order we write in the text
``$\overrightarrow{S}$,'' this should be viewed as being replaced by 
the string ``$s_1 < s_2 < \cdots < s_{||S||}$.''
The voter set $\Hat{V}$ consists of the two subparts---voters
simulating voters from the underlying elections, and 
voters who are ``normalizing'' voters.
The total number of voters will be
$(2\sum_j ||V_j||)-1$, which is odd as required by the
statement of the lemma being proven.
We now 
describe the simulating voters (the cases of $1$
and $k$ are exactly analogous to the other cases, but are 
stated separately just for notational reasons):
\begin{itemize}
\item There will be voters simulating 
the voters of $V_1$.  In particular, 
for each voter $\voter{e_1<e_2<\cdots<e_{||C_1||}}$ in $V_1$, we create
a voter
\[\voter{ \overrightarrow{S} < 
\overrightarrow{C_2} < \cdots  
< \overrightarrow{C_k} < 
e_1<e_2<\cdots<e_{||C_1||}}.\]
Note that $c$ is one of the $e_j$'s.
\item For each $i$, $1 < i < k$, there will be voters simulating 
the voters of $V_i$.  In particular, for each $i$, $1<i<k$, and 
for each voter $\voter{e_1<e_2<\cdots<e_{||C_i||}}$ in $V_i$, we create
a voter
\[\voter{ \overrightarrow{S} < 
\overrightarrow{C_1} < \cdots  < \overrightarrow{C_{i-1}} < 
\overrightarrow{C_{i+1}} < \cdots < \overrightarrow{C_k} < 
e_1<e_2<\cdots<e_{||C_i||}}.\]
Note that $c$ is one of the $e_j$'s.
\item There will be voters simulating 
the voters of $V_k$.  In particular, 
for each voter $\voter{e_1<e_2<\cdots<e_{||C_k||}}$ in $V_k$, we create
a voter
\[\voter{ \overrightarrow{S} < 
\overrightarrow{C_1} < \cdots 
< \overrightarrow{C_{k-1}} < 
e_1<e_2<\cdots<e_{||C_k||}}.\]
Note that $c$ is one of the $e_j$'s.
\end{itemize}
For each $i$,
we want $c$'s behavior with respect to candidates in $C_i$ to depend only
on voters that simulate~$V_i$. That is, every candidate in $C_i\setminus \{c\}$
should be preferred to $c$ by exactly half of the voters in $\Hat{V}$ that do
{\em not\/} simulate $V_i$.  
To accomplish this, we add  $(\sum_j ||V_j||) - 1$  
normalizing voters.  
\begin{itemize}
\item There will be $(\sum_j ||V_j||) - 1$  normalizing voters.
Each normalizing voter will have preferences of the form
$$ \mbox{``some of the $\overrightarrow{C_j}$'s''} 
< c < \overrightarrow{S} < 
\mbox{``the rest of the $\overrightarrow{C_j}$'s.''}$$
Within the ``some of'' and ``rest of'' blocks, the order of the 
candidates can be 
arbitrary.
So all that remains to do is to specify, for each particular
one of the normalizing voters, how to decide which 
$\overrightarrow{C_j}$'s go to the left of $c$ (the ``some
of'' block), and which go to the 
right of $\overrightarrow{S}$ (the ``rest of'' 
block).  Let us do so.  Let the normalizing 
voters be named $\sigma_1$,~$\ldots$, $\sigma_{(\sum_j ||V_j||)-1}$.
Consider normalizing voter $\sigma_q$.  Then, for each $i$, 
in the preference of $\sigma_q$ let it be the case
that $\overrightarrow{C_i}$ goes to the right of 
$\overrightarrow{S}$ if
$$q 
\leq \lfloor ||V_i||/2 \rfloor + \sum_{j\neq i} ||V_j||,$$ 
and otherwise $\overrightarrow{C_i}$ goes to the left of $c$.
Note that, for each $i$, exactly 
$ \lfloor ||V_i||/2 \rfloor + \sum_{j\neq i} ||V_j||$ normalizing
voters will have $\overrightarrow{C_i}$ to the right of $S$
and exactly $\lfloor ||V_i||/2 \rfloor$ normalizing voters 
will have $\overrightarrow{C_i}$ to the left of $c$.
\end{itemize}

Recall that $c = c_1 = \cdots = c_k$.  We have to prove that 
$ \sum_j \score(\pair{C_j,c,V_j}) =
\score(   
\pair{\Hat{C},c,\Hat{V}})$.

First note that 
$c$ is preferred to each candidate in $S$ by 
$\sum_j ||V_j||$ of the $(2\sum_j ||V||) - 1$ voters in~$\Hat{V}$.
Also, for each $i$, 
it holds that $c$ is preferred to all candidates in $C_i\setminus \{c\}$
by exactly half of the voters that do {\em not\/} simulate $V_i$.  To
see this, note 
that 
$c$ is preferred to each candidate in $C_i\setminus \{c\}$ by all voters that
simulate
a $V_j$ with $j\neq i$, and is  
also preferred by $\lfloor ||V_i||/2 
\rfloor$ of the normalizing voters. Thus,
$c$ is preferred to each candidate in $C_i \setminus \{c\}$
by $(\sum_{j \neq i} ||V_j||) + \lfloor ||V_i||/2\rfloor)$ 
of the $(\sum_{j \neq i} ||V_j||) + (\sum_j ||V_j||) - 1$ voters
not simulating~$V_i$, which indeed is exactly half of 
the voters not simulating~$V_i$ (recall that $||V_i||$ is odd).

For each $i$, let $K_i = \score(\pair{C_i,c,V_i})$.
Then after $K_i$ 
switches in~$V_i$, $c$ is preferred to $e$ by more than $||V_i||/2$
voters in $V_i$, for each $e \in C_i\setminus \{c\}$. 
This implies that after 
the analogous $K_i$ switches in $\Hat{V}$ (i.e., in the 
voters in $\Hat{V}$ that 
simulate $V_i$), $c$ is preferred to $e$
by more than $||V_i||/2$ voters in
that part of $\Hat{V}$ that simulates~$V_i$, for each $e \in
C_i\setminus \{c\}$, and thus by more than half of the voters in $\Hat{V}$. 
It follows that $\sum_j K_j $ 
switches in voters of $\Hat{V}$ turn $c$ into a Condorcet winner.
This proves that $\score(\pair{\Hat{C},c,\Hat{V}}) \leq 
\sum_j \score(\pair{C_j,c,V_j})$.

It remains to show that $\score(\pair{\Hat{C},c,\Hat{V}}) \geq
\sum_j \score(\pair{C_j,c,V_j})$.
Let $\Hat{K} = \score(\pair{\Hat{C},c,\Hat{V}})$. 
Then $\Hat{K}$ switches in
$\Hat{V}$ turn $c$ into a Condorcet winner. 
If \mbox{$\Hat{K} \geq ||S||$}, then
$\Hat{K} > 
\sum_j \score(\pair{C_j,c,V_j})$,
since $\score(\pair{C_j,c,V_j}) \leq ||V_j||\cdot (||C_j||-1)$
and so
$\sum_j \score(\pair{C_j,c,V_j}) \leq 
\sum_j ||V_j||\cdot (||C_j||-1) < 
\sum_j ||V_j||\cdot ||C_j|| = ||S||$. 
So \mbox{$\Hat{K} \geq ||S||$} is impossible,
and we thus know that $\Hat{K} < ||S||$. With less
than $||S||$ switches, $c$ cannot gain extra 
votes over candidates in $(\bigcup_j C_j) \setminus
\{c\}$ in normalizing voters, as can be immediately seen
in light of the preferences of the 
normalizing voters.  Also, for each $i$:
Since $c$ is already preferred to all
candidates in 
$C_i\setminus \{c\}$ by all voters that simulate~$V_j$ with $j \neq i$,
$c$ cannot gain extra votes over 
candidates in $C_i\setminus \{c\}$ in voters simulating $V_j$ with $j \neq i$.
It follows that $c$ can gain extra 
votes over candidates in $C_i\setminus \{c\}$ only in voters
that simulate~$V_i$. After $\Hat{K}$ switches, $c$ is still preferred to all
candidates in $C_i \setminus \{c\}$ by at most half of the voters that do not
simulate $V_i$,
and 
at the same time, $c$ has become a Condorcet winner.
It follows that after these $\Hat{K}$ switches,  $c$ is
preferred to $e$ by more than $||V_i||/2$ of the voters that simulate $V_i$, 
for each $e$ in $C_i\setminus \{c\}$. Let $M_i$ be the number of
switches that take place in the voters of 
$\Hat{V}$ that simulate~$V_i$. Then
$M_i \geq \score(\pair{C_i,c,V_i})$.

Since this argument applies for all $i$, 
it follows that 
$$\score(\pair{\Hat{C},c,\Hat{V}}) = \Hat{K} \geq
\sum_j M_j  \geq
\sum_j \score(\pair{C_j,c,V_j}),$$
proving the lemma.~\qed

\medskip

\noindent {\bf Proof of Lemma~\ref{l:less}.} 
Let $A$  and $f$ be the NP-complete set and the reduction from
Lemma~\ref{l:reduction}, and let $\dsum$ be the function from
Lemma~\ref{l:SUMMATION}.  We seek to apply Lemma~\ref{l:klaus},
using the $A$ (i.e., {\tt{}3DM}) of Lemma~\ref{l:reduction} 
as the $A$ of Lemma~\ref{l:klaus}, using {\tt{}2ER} as the 
$B$ of Lemma~\ref{l:klaus}, and using a function $g$ that 
we will define in this proof as the $g$ of Lemma~\ref{l:klaus}.

Let $x_1,\ldots,x_{2k} \in \sigmastar$ be such that $\chi_A(x_1) \geq
\cdots \geq \chi_A(x_{2k})$.
For $i = 1, \ldots, 2k$, let $f(x_i) = \pair{\pair{C_i, c_i, V_i}, K_i}$.
We will write $S_i$ for the Dodgson triple 
$\pair{C_i,c_i,V_i}$.
We will compare the Dodgson score of the sum of the even Dodgson triples
with the Dodgson score of the sum of the odd Dodgson triples, i.e., we
will look at the value of
\[\score(
\dsum ( \pair{ S_{2}, S_4, \ldots, S_{2k} } ) 
)  -
\score(\dsum ( \pair{ S_{1}, S_3, \ldots, S_{2k-1} } ) ).\]
By Lemma~\ref{l:SUMMATION}, this is the same as
\[\sum_{1\leq i\leq k} (\score(S_{2i}) -
\score(S_{2i-1})).\]

Recall that $\chi_A(x_1) \geq
\cdots \geq \chi_A(x_{2k})$.
If $||\{i \condition  x_i \in A\}||$ is even then, for all $i$, $1\leq i
\leq k$, it holds that $x_{2i -1}
\in A  \Longleftrightarrow  x_{2i} \in A$.
So, by Lemma~\ref{l:reduction}, for each~$i$, 
either $\score(S_{2i-1}) = K_{2i-1}$
and $\score(S_{2i}) = K_{2i}$, or
$\score(S_{2i-1}) = K_{2i-1}+1$
and $\score(S_{2i})= K_{2i}+1$.
It follows that, for each~$i$, $1\leq i \leq k$,
$$\score(S_{2i}) -  \score(S_{2i-1}) = K_{2i} -  K_{2i-1}.$$

On the other hand, if 
$||\{i \condition  x_i \in A\}||$ is odd then, for some $j$, $1\leq j
\leq k$,
$x_{2j -1} \in A$ and  $x_{2j}  \not \in A$ and,
for all $i\neq j$, $1\leq i \leq k$, 
it holds that $x_{2i -1} \in A 
\Longleftrightarrow x_{2i} \in A$.
It follows that 
$\score(S_{2j}) -
\score(S_{2j-1}) = 1 + K_{2j} -  K_{2j-1}$ and,
for all $i \neq j$, $1\leq i \leq k$,
$\score(S_{2i}) -
\score(S_{2i-1}) = K_{2i} -  K_{2i-1}$.

To summarize, 
\begin{eqnarray*}
\lefteqn{\score(
\dsum ( \pair{ S_{2}, S_4, \ldots, S_{2k} } ) 
) -
\score(
\dsum ( \pair{ S_{1}, S_3, \ldots, S_{2k-1} } ) 
) =} \\
 & & \left\{ \begin{array}{ll}
\sum_{1\leq i\leq k} K_{2i} - \sum_{1\leq i\leq k} K_{2i-1} &
\mbox{if $||\{i \condition  x_i \in A\}||$ is even, and}\\
1 + \sum_{1\leq i\leq k} K_{2i} - \sum_{1\leq i\leq k} K_{2i-1}&
\mbox{if $||\{i \condition  x_i \in A\}||$ is odd.}
\end{array}
\right.
\end{eqnarray*}

This implies that $||\{i \condition  x_i \in A\}||$ is odd if and only if
\[\score(
\dsum ( \pair{ S_{2}, S_4, \ldots, S_{2k} } ) 
) + \sum_{1 \leq i \leq k} K_{2i-1} \geq \]
\[\score(
\dsum ( \pair{ S_{1}, S_3, \ldots, S_{2k-1} } ) 
)  + 1 + \sum_{1 \leq i \leq k} K_{2i}. \]

For any integer $m \geq 1$, define a Dodgson triple 
\[
T_{m} = \pair{\{i \condition 1\leq i \leq m+1\},
1, \{\voter{1 < 2 < 3 < \cdots < m+1}\}}.
\]
Then $T_m$ has an odd number of voters (namely one), and $\score(T_m) = m$. 
Thus, again by Lemma~\ref{l:SUMMATION}, 
$||\{i \condition  x_i \in A\}||$ is odd if and only if
\[\score(
\dsum ( \pair{ S_{2}, S_4, \ldots, S_{2k} ,  
T_{\sum_{1 \leq i \leq k} K_{2i-1}} } ))
\geq \]
\[\score(
\dsum ( \pair{ S_{1}, S_3, \ldots, S_{2k-1}, 
T_{1+\sum_{1 \leq i \leq k} K_{2i}} } )
).\]

Given $x_1,\ldots,x_{2k}$, define the function $g(x_1,\ldots,x_{2k}) =
\pair{\pair{C,c,V},\pair{D,d,W}}$, 
where 
$$ \pair{C,c,V} = \dsum ( \pair{ S_{1}, S_3, \ldots, S_{2k-1}, 
T_{1+\sum_{1 \leq i \leq k} K_{2i}} } )$$
and 
$$ \pair{D,d,W} = 
\dsum ( \pair{ S_{2}, S_4, \ldots, S_{2k} ,  
T_{\sum_{1 \leq i \leq k} K_{2i-1}} } ),$$
and 
(without loss of generality, via trivial renaming if 
necessary) $c
\neq d$.

Note that $g(x_1,\ldots,x_{2k})$ is computable in time 
polynomial in $|x_1| + |x_2| + \cdots + |x_{2k}| + 2k - 1$
(recall the conventions regarding variable-arity functions 
discussed in Section~\ref{s:prelims} and 
footnote~\ref{f:arity}).
Since
\[
\score(\pair{C,c,V}) \leq \score(\pair{D,d,W}) \Longleftrightarrow
||\{i \condition  x_i \in A\}|| \mbox{ is odd},
\]
it follows by Lemma~\ref{l:klaus} that the problem {\tt{}2ER} 
is $\thetatwo$-hard.~\qed

\medskip

\noindent {\bf Proof of Lemma~\ref{l:merge}.} 
Without loss of generality, we assume that $||V|| \geq ||W||$
and that \mbox{$C \cap D = \emptyset$}.
Also, enumerate $C \setminus \{c\}$ as $\{c_1,c_2, \ldots, c_{||C||-1}\}$, and
$D \setminus \{d\}$ as $\{d_1,d_2, \ldots, d_{||D||-1}\}$.

The construction and proof are similar in flavor 
to the construction and proof of
Lemma~\ref{l:SUMMATION}. 
However, in this proof, the 
number of voters has to be even, 
as we seek to ensure
that $c$
is preferred to $d$ by exactly half of the voters.

We define 
a set of ``separating'' candidates:
$S = \{s_i \condition 1 \leq i \leq 2(||C||\cdot
||V|| + ||D||\cdot ||W||)\}$.
We will also use another set of separating candidates,
$T = \{ t_i \condition 1 \leq i \leq ||S||\}$, 
of the same cardinality as~$S$. Let $m = ||S||/2$. 
Let $\Hat{C} = C \cup D \cup S \cup T$.
The set of new voters $\Hat{V}$ consists of the following subparts:
\begin{description}
\item[(a)] Voters simulating $V$:
for each voter $\voter{e_1<e_2<\cdots<e_{||C||}}$ in $V$, we create
a voter
\[
\hspace*{-8mm}
\voter{d < s_1<\cdots<s_{||S||}<d_1<\cdots<d_{||D||-1}<
t_1<\cdots<t_{||T||}<e_1<e_2<\cdots<e_{||C||}}.\]
\item[(b)] Voters simulating $W$:
for each voter $\voter{e_1<e_2<\cdots<e_{||D||}}$ in $W$, we create a
voter
\[
\hspace*{-8mm}
\voter{t_1<\cdots<t_{||T||}<c < s_1<\cdots<s_{||S||}<c_1<\cdots<c_{||C||-1}<
e_1<e_2<\cdots<e_{||D||}}.\]
\end{description}
In addition, we create $||V||+1$ normalizing voters (recall
that $||V||$ and $||W||$ are both odd), consisting of three subparts:
\begin{description}
\item[(c)]
$\lceil ||V||/2 \rceil - \lceil ||W||/2 \rceil$ voters:
\[
\hspace*{-8mm}
\voter{t_1<\cdots<t_{||T||}<c < s_1 < \cdots < s_{||S||} < c_1 < \cdots < c_{||C||-1} <
d_1 < \cdots < d_{||D||-1} < d}.\]
\item[(d)]
$\left\lceil ||V||/2 \right\rceil$ voters:
\[
\hspace*{-8mm}
\voter{
t_1<\cdots<t_{||T||}<c_1 < \cdots < c_{||C||-1} < d_1 < \cdots <
d_{||D||-1} < s_{||S||} < \cdots < s_1 < c < d}.\]
\item[(e)]
$\left\lceil ||W||/2 \right\rceil$ voters:
\[
\hspace*{-8mm}
\voter{
t_1<\cdots<t_{||T||}< c_1 < \cdots < c_{||C||-1} < d_1 < \cdots <
d_{||D||-1}< s_1 < \cdots < s_{||S||} < d < c}.\]
\end{description}

The above construction of $\Hat{C}$ and $\Hat{V}$ defines our
functions $\merge(\pair{C,c,V}, \pair{D,d,W}) =
\pair{\widehat{C},c,d,\widehat{V}}$ and $\mergeprime(\pair{C,c,V},
\pair{D,d,W}) = \pair{\widehat{C},c,\widehat{V}}$.  These
functions clearly satisfy
properties (i) and (ii) of Lemma~\ref{l:merge}.

To satisfy properties (iii) and (iv), we have to prove that 
$\score(\pair{\Hat{C},c,\Hat{V}}) = 
\score(\pair{C,c,V}) + 1$ and that $\score(\pair{\Hat{C},d,\Hat{V}})
=\score(\pair{D,d,W}) + 1$.

First note
that $c$ is preferred to every candidate 
in $S \cup D \setminus \{d\}$ by 
$||V|| + \left\lceil ||V||/2 \right\rceil + 
\left\lceil ||W||/2 \right\rceil$ of
the $2||V|| + ||W|| + 1$ voters in $\Hat{V}$.
Similarly, 
$d$ is preferred to every candidate in $S \cup C \setminus \{c\}$ by 
$||W|| + ||V|| + 1$ of the $2||V|| + ||W|| + 1$ voters in $\Hat{V}$.
Similarly, 
$c$ is preferred to 
each $t\in T$ by all voters in $\Hat{V}$,
and $d$ is preferred to each $t\in T$
by
$||V|| + ||W|| + 1$ of the $2||V|| + ||W|| + 1$ voters in $\Hat{V}$.

In addition, $c$  is preferred to all candidates in $C\setminus \{c\}$
by $\left\lceil ||V||/2 \right\rceil + \left\lceil ||W||/2 \right\rceil =
(||V||+||W||)/2 + 1$ of the $||V||+||W|| + 1$ voters that do {\em not\/} simulate
$V$. Likewise, $d$ is preferred to
all candidates in $D\setminus \{d\}$
by $||V|| + 1$ of the $2||V||+1$ voters not simulating $W$.
Finally, $c$ is preferred to $d$ by $||V|| + \left\lceil ||W||/2 \right\rceil =
(2||V||+||W||+1)/2$ of the $2||V||+||W||+1$ voters in $\Hat{V}$---exactly
half.

Let $K = \score(\pair{C,c,V})$.
Then after  $K$ switches in $\Hat{V}$, $c$ is preferred to $e$
by more than $||V||/2$ voters in
that part of $\Hat{V}$ that simulates $V$, for every \mbox{$e \in
C\setminus \{c\}$}, and 
thus by more than half of the voters in $\Hat{V}$. 
It follows that after $K$ switches, $c$ is preferred to $e$ by a majority of
voters, for all $e \in \Hat{C} \setminus \{c,d\}$.
If, in addition to these $K$ switches, we switch $c$ and $d$ in a normalizing
voter of the form 
$\voter{t_1 < \cdots < t_{||T||} < c_1 < \cdots < c_{||C||-1} < d_1 < \cdots <
d_{||D||-1} < s_{||S||} < \cdots < s_1 < c < d}$,  
then $c$ has become a Condorcet winner.
Thus $\score(\pair{\Hat{C},c,\Hat{V}}) \leq K + 1 = \score(C,c,V) + 1$.
In exactly the same way, we can show that  $\score(\pair{\Hat{C},d,\Hat{V}}) \leq
\score(\pair{D,d,W} ) + 1$.

It remains to show that $\score(\pair{\Hat{C},c,\Hat{V}}) \geq
\score(\pair{C,c,V}) + 1$ and that  $\score(\pair{\Hat{C},d,\Hat{V}}) \geq
\score(\pair{D,d,W}) + 1$.
Let $\Hat{K} = \score(\pair{\Hat{C},c,\Hat{V}})$. Then $\Hat{K}$ switches in
$\Hat{V}$ turn $c$ into a Condorcet winner. 
Recall that $m=||S||/2$.  If $\Hat{K} \geq m$, then
$\Hat{K} > \score(\pair{C,c,V}) + 1$, since 
$\score(\pair{C,c,V}) < ||C|| \cdot ||V|| < m$ (recall $||W||$
is odd and thus nonzero, and without loss of generality
we assume $||D||>0$).
So $\Hat{K} \geq m$ is impossible, which implies that $\Hat{K} < m$.
In order to become a Condorcet winner, $c$ in particular 
needs to gain one vote over~$d$. 
With less than $m$ switches, the {\em only\/} way in which $c$ can gain
this vote is by switching $c$ and $d$ in a normalizing voter of the form
$\voter{t_1 < \cdots < t_{||T||} < c_1 < \cdots < c_{||C||-1} < d_1 < \cdots <
d_{||D||-1} < s_{||S||} < \cdots < s_1 < c < d}$.
This uses one of the $\Hat{K}$ switches.

With less than $m$ switches, $c$ cannot 
gain extra votes over candidates in $C \setminus
\{c\}$ in normalizing voters, or in voters that simulate~$W$.
It follows that $c$ can gain extra votes over candidates in $C\setminus \{c\}$ only in voters
that simulate~$V$. After $\Hat{K}$ switches, $c$ is still preferred to all
candidates in $C\setminus \{c\}$ by at most the smallest possible majority of
the (odd) number of voters that do not simulate $V$, and at the
same time,
$c$ has become a Condorcet winner. Since $||V||$ is odd, it follows that, after
these
$\Hat{K}$ switches,  $c$ is preferred to $e$ by more than $||V||/2$ voters that
simulate
$V$, for every $e$ in $C\setminus \{c\}$.
Let $\Hat{K}_{V}$ be the number of
switches that take place in the voters of $\Hat{V}$ that simulate~$V$. Then
$\Hat{K}_{V} \geq \score(\pair{C,c,V})$.
Since we had to use one switch to switch $c$ and $d$ in a normalizing voter,
$$\score(\pair{\Hat{C},c,\Hat{V}}) = \Hat{K} \geq \Hat{K}_{V}+ 1 \geq
\score(\pair{C,c,V}) + 1.$$

The same argument can be used to show that 
$$\score(\pair{\Hat{C},d,\Hat{V}}) \geq
\score(\pair{D,d,W}) + 1,$$
which proves properties (iii) and~(iv).

Finally, we prove property (v) of the lemma: For
each \mbox{$e \in \widehat{C}\setminus \{c,d\}$},
$\score(\pair{\widehat{C},c,\widehat{V}}) <
\score(\pair{\widehat{C},e,\widehat{V}})$.  First note that we have
chosen $S$ sufficiently large to ensure that
$\score(\pair{\widehat{C},c,\widehat{V}}) < m$, since
\mbox{$\score(\pair{C,c,V}) < ||C||\cdot ||V|| < m$} and 
$\score(\pair{\widehat{C},c,\widehat{V}}) = 
\score(\pair{C,c,V}) + 1$ by property~(iii).

Consider $t_{||T||}$. In order to become a Condorcet winner,
$t_{||T||}$ must in particular outpoll $d$ in pairwise elections.
In the specified preferences, $t_{||T||}$ is preferred to $d$
by $||V||$ of the $2||V|| + ||W|| + 1$ voters 
in $\widehat{V}$.
Thus, more than $\left\lceil ||W||/2 \right\rceil$ of the voters not
simulating $V$ must be convinced to prefer $t_{||T||}$ to~$d$. 
However,
to gain even one additional vote over~$d$ amongst 
the voter groups~(b), (c), (d), and~(e),
$t_{||T||}$ would require more than $m$ switches upwards. Since
$\score(\pair{\widehat{C},c,\widehat{V}}) < m$, the score of $c$
is less than that of~$t_{||T||}$. The same argument applies to any
$t_i$, $1 \leq i \leq ||T||$.

Consider $s_{||S||}$. In order to become a Condorcet winner,
$s_{||S||}$ must in particular outpoll $c$ in pairwise elections.
Initially, $s_{||S||}$ is preferred to $c$ by \mbox{$||W|| +
\lceil ||V||/2 \rceil - \lceil ||W||/2 \rceil 
= \left\lceil ||V||/2 \right\rceil + \left\lceil
  ||W||/2 \right\rceil - 1$} voters,
namely those belonging to (b) and~(c).  Thus,
more than $\left\lceil ||V||/2 \right\rceil$ of the voters
amongst~(a),~(d), and~(e) must be convinced to prefer $s_{||S||}$ to~$c$.
However, to gain one more vote over~$c$ in~(a), $s_{||S||}$ would need more
than $m$ switches upwards. Likewise, for $s_{||S||}$ to gain one
more vote over~$c$ in~(d), it would also have to switch more than
$m$ times upwards. Finally, to gain one more vote over~$c$ in~(e),
$s_{||S||}$ needs only two switches per vote. However, since there are no
more than \mbox{$\left\lceil ||W||/2 \right\rceil \leq \left\lceil
  ||V||/2 \right\rceil$} voters in~(e) and $s_{||S||}$ needs to be
preferred over $c$ by {\em more\/} than $\left\lceil ||V||/2
\right\rceil$ additional voters, $s_{||S||}$ cannot become a Condorcet
winner by changing only the minds of the voters in~(e).  It follows
that \mbox{$\score(\pair{\Hat{C},c,\Hat{V}}) <
  \score(\pair{\Hat{C},s_{||S||},\Hat{V}}) $}.

Consider $s_1$.  As above, for $s_1$ to become a Condorcet winner,
more than $\left\lceil ||V||/2 \right\rceil$ of the voters
amongst~(a),~(d), and~(e) must be convinced to prefer $s_1$ to~$c$ in
particular. Now, to gain one vote in either (a) or (e) requires more
than $m$ switches. However, similarly
to the previous paragraph,  the remaining $\left\lceil ||V||/2
\right\rceil$ voters in (d) alone are too few to make $s_1$ a
Condorcet winner. It follows that
\mbox{$\score(\pair{\Hat{C},c,\Hat{V}}) <
  \score(\pair{\Hat{C},s_1,\Hat{V}}) $}. Note that for each $s \in
S\setminus \{s_1,s_{||S||}\}$, at least one of the two given arguments
(the one for $s_{||S||}$ and the one for~$s_1$) apply, yielding
\mbox{$\score(\pair{\Hat{C},c,\Hat{V}}) <
  \score(\pair{\Hat{C},s,\Hat{V}}) $}, since each such $s$ needs more
than $m$ switches (because $m=||S||/2$) to gain one vote in either (d) or~(e).

Finally, consider $d_{||D||-1}$. As was the case for the elements
of~$S$, more than $\left\lceil ||V||/2 \right\rceil$ of the voters
amongst~(a),~(d), and~(e) must be convinced to prefer $d_{||D||-1}$ to $c$
in order for $d_{||D||-1}$ to become a Condorcet winner. However, more than
$m$ switches would be required to gain even one vote from one
of~(a),~(d), or~(e). Thus \mbox{$\score(\pair{\Hat{C},c,\Hat{V}}) <
  \score(\pair{\Hat{C},d_{||D||-1},\Hat{V}}) $}. The same argument
applies to each element in $(C \cup D) \setminus \{c,d\}$.  To
summarize, we have shown that property~(v) holds.~\qed

\medskip

Having proven these lemmas, we may now turn to 
the proof of the Theorem~\ref{t:theta}.

\medskip

\noindent {\bf Proof of Theorem~\ref{t:theta}.} 
To prove the $\thetatwo$-completeness of \dr, we must according to the
definition prove both an upper bound ($\dr\in\thetatwo$)
and a lower bound ($\dr$ is $\thetatwo$-hard).  

To prove the lower bound, it suffices to
provide
a 
$\leq_{m}^{p}$-reduction, $f$, from {\tt 2ER} to \dr.  
$f$ is defined as follows.  Let $t_o$ be some fixed
string that is not in \dr.  $f(x)$ is defined as 
being $t_o$ if $x$ is not an instance of {\tt{}2ER},
and as being $\merge(x_1,x_2)$ otherwise, where 
$x=\pair{x_1,x_2}$ and
$\merge$ is as defined in Lemma~\ref{l:merge}.
Note that $\merge$ and thus $f$ are
polynomial-time computable.  Note also
that for any instance
$\pair{\pair{C,c,V},\pair{D,d,W}}$ of {\tt 2ER}, it holds that if
$\pair{\Hat{C},c,d,\Hat{V}} = \merge(\pair{C,c,V},\pair{D,d,W})$, then
\[
\pair{\pair{C,c,V},\pair{D,d,W}} \in \mbox{\tt 2ER} \ 
\Longleftrightarrow\ 
\pair{\Hat{C},c,d,\Hat{V}} \in \dr
\]
by properties (iii) and (iv) of Lemma~\ref{l:merge}.
Note also that for any input $x$ that is not an instance
of {\tt{}2ER}, $f(x)$ maps to $t_o$, a string
that is not in \dr.
Thus, $f$ is a $\leq_m^p$-reduction 
from {\tt 2ER} to \dr.  
From
Lemma~\ref{l:less}, $\thetatwo$-hardness of \dr\ follows immediately.

 Finally, we claim that \dr\ is in~$\thetatwo$.  
 This can be seen as follows.
 We can in parallel 
 ask all plausible \ds\ queries for each of the two designated
 candidates, say $c$ and~$d$, and from this compute the exact score of each 
 of $c$ and $d$ and thus we can tell whether $c$ ties-or-defeats~$d$.
 Note that there is a polynomial upper bound on the highest
 possible score (this is what was meant above
 by ``plausible''), and thus this procedure indeed can be 
implemented via a polynomial-time truth-table reduction to the
NP-complete set \ds.  However, 
the class of languages accepted via polynomial-time
truth-table reductions to NP sets coincides 
with $\thetatwo$~\cite{hem:j:sky,koe-sch-wag:j:diff}.
This establishes the upper bound, i.e., 
that $\dr \in \thetatwo$.~\qed

\medskip

\dw\ is similarly $\thetatwo$-complete.

\begin{theorem}\label{t:dodgson-winner-complete}
\dw\ is  $\thetatwo$-complete.
\end{theorem}
\begin{corollary}
If \dw\ is $\np$-complete, then $\ph = \np$.
\end{corollary}

Bartholdi et al.~\cite{bar-tov-tri:j:who-won}
have stated without proof that $\dr
\leq_m^p \dw$.  Theorem~\ref{t:theta} plus this 
assertion 
would prove Theorem~\ref{t:dodgson-winner-complete}.  However, as we wish
our proof to be complete, we now prove
Theorem~\ref{t:dodgson-winner-complete}.  (We note in passing
that our paper implicitly 
provides an indirect proof of their assertion.  In particular, 
given that one
has proven Theorem~\ref{t:theta} and 
Theorem~\ref{t:dodgson-winner-complete}, the assertion follows, 
since it follows 
from the definition of $\thetatwo$-completeness that 
all $\thetatwo$-complete problems are $\leq_m^p$-interreducible.)

\medskip

\noindent {\bf Proof of Theorem~\ref{t:dodgson-winner-complete}.} 
As in the case of \dr, $\dw \in \thetatwo$ is easily seen to hold,
since we can in parallel ask all plausible \ds\ queries for each of
the given candidates (note that the number of 
candidates and the highest possible score for each
candidate are both 
polynomially bounded in the input length) and thus can compute the
exact Dodgson score for each candidate. After having
done so, it is easy to decide
whether or not the designated candidate $c$ ties-or-defeats all other
candidates in the election. This proves the upper bound.

To prove the lower bound, we will provide a polynomial-time
many-one reduction from {\tt 2ER} to \dw.  By Lemma~\ref{l:less},
the claim of this theorem then follows.
In fact, the following function $f$ 
provides
a polynomial-time
many-one reduction from {\tt 2ER} to \dw.  
Let $t_o$ be some fixed
string that is not in \dw.  $f(x)$ is defined as 
being $t_o$ if $x$ is not an instance of {\tt{}2ER},
and as being $\mergeprime(x_1,x_2)$ otherwise, where 
$x=\pair{x_1,x_2}$ and 
$\mergeprime$ is as defined in Lemma~\ref{l:merge}.
To see that this is correct, note that $f$ is 
polynomial-time computable, and that when $x$ is not 
an instance of {\tt{}2ER} then $f(x)$ is not 
in \dw.  

We now turn to the behavior of $f(x)$ when $x$ is 
an instance of {\tt{}2ER}.  
Given any 
pair of Dodgson triples, $\pair{C,c,V}$ and $\pair{D,d,W}$,
for which both $||V||$ and $||W||$ are odd and $c \neq d$, let
$\pair{\widehat{C},c,\widehat{V}} = \mergeprime(\pair{C,c,V},
\pair{D,d,W})$. Assume $\score(\pair{C,c,V}) \leq
\score(\pair{D,d,W})$. By properties (iii) and~(iv)
of Lemma~\ref{l:merge}, it follows that
$\score(\pair{\Hat{C},c,\Hat{V}}) \leq
\score(\pair{\Hat{C},d,\Hat{V}})$ as well.  However, since by property~(v)
of Lemma~\ref{l:merge}
$\score(\pair{\Hat{C},c,\Hat{V}}) < \score(\pair{\Hat{C},e,\Hat{V}})$
for every $e \in \Hat{C}\setminus \{c,d\}$, it follows that $c$ is a
winner of the election specified by $\Hat{C}$ and 
$\Hat{V}$.
Conversely, assume $\score(\pair{C,c,V}) > \score(\pair{D,d,W})$.
Again, properties (iii) and (iv) of Lemma~\ref{l:merge} imply that
$\score(\pair{\Hat{C},c,\Hat{V}}) > \score(\pair{\Hat{C},d,\Hat{V}})$.
Thus, $c$ is not a winner of the election specified by $\Hat{C}$ and 
$\Hat{V}$.~\qed

\medskip

Finally,
recall that our multisets are specified as a list containing, for 
each voter, the preference order of that voter.  Our main
theorem, Theorem~\ref{t:dodgson-winner-complete},
proves that checking if a candidate is a Dodgson
winner is $\thetatwo$-complete.  
Is this complexity coming from the number of candidates, or is the
problem already complex with, for example, fixed numbers of candidates?
In fact, for each fixed constant $k$, there clearly is a 
polynomial-time algorithm to compute (all) Dodgson scores, and thus
all Dodgson winners, in elections having at most $k$ candidates.

\begin{proposition} {\bf \cite{bar-tov-tri:j:who-won}}\label{p:algorithm}
Let $k$ be any fixed positive integer.  There is a
polynomial-time algorithm $A_k$ that computes all Dodgson scores (and
thus all Dodgson winners) in Dodgson elections having at most $k$
candidates.  
\end{proposition}

Proposition~\ref{p:algorithm} in no way conflicts with
Theorem~\ref{t:dodgson-winner-complete}.  In fact, though each $A_k$
is a polynomial-time algorithm, the degree of the polynomial runtimes
of the $A_k$ is itself exponential in~$k$.  It is also known
that, for each fixed constant $k$, there is a polynomial-time
algorithm to compute all Dodgson winners in elections having at most
$k$ voters~\cite{bar-tov-tri:j:who-won}.

\section{Conclusions}
This paper establishes that testing whether a given candidate wins
a Dodgson election is $\thetatwo$-complete, thus providing the 
first truly natural complete problem for the class~$\thetatwo$.

In this paper, we assumed that no voter views any two candidates as
being of equal desirability.  However, note that if one allows such
ties, our $\thetatwo$-hardness result remains valid, as our case is
simply a special case of this broader problem.  On the other hand, it
is not hard to see that the broader problem remains in~$\thetatwo$ (in
both of 
the natural models of switches involving ties, i.e., the model in
which moving from $\langle a = b < c \rangle$ to $\langle c < a = b
\rangle$ requires just one switch, and the model in which 
this requires
two separate switches).  Thus, this broader problem is also
$\thetatwo$-complete.

Since this paper first appeared, some related work has been 
done that may be of interest to readers of this paper.  
Hemaspaandra and Wechsung~\cite{hem-wec:ctoappear:mee}
have shown that the minimization problem for boolean 
formulas is $\thetatwo$-hard;  it remains open
whether that problem is $\thetatwo$-complete.  
Hemaspaandra and Rothe~\cite{hem-rot:t:max-independent-set-by-greedy}
have shown that 
recognizing the instances on which the greedy
algorithm can obtain independent sets that are within a certain
fixed factor of optimality is itself a $\thetatwo$-complete 
task.  Hemaspaandra, Hemaspaandra, and Rothe
have discussed the relationship between raising 
a problem's lower bound
from NP-hardness to $\thetatwo$-hardness and its potential
solvability via such modes of computation as randomized
and approximate computing~\cite{hem-hem-rot:j:raising-lower-bounds}.

{\samepage
\begin{center}
{\bf Acknowledgments}
\end{center}
\nopagebreak
\indent 
We are indebted to J.~Banks and R.~Calvert for recommending
Dodgson elections to us as an interesting open topic worthy of study,
and for providing us with the literature on this topic.  
We are grateful to M.~Scott for suggesting to us the study of 
the fixed-parameter complexity of Dodgson scoring. 
We
thank D.~Austen-Smith, J.~Banks,
R.~Calvert, A.~Rutten, and J.~Seiferas for helpful conversations and
suggestions.  
L.~Hemaspaandra thanks J.~Banks and R.~Calvert for
arranging, and J.~Banks for supervising, his Bridging Fellowship at
the University of Rochester's Department of Political Science, during
which this project was started.  

}%

{\singlespacing

\bibliography{../../gry}

}

\end{document}